\documentclass[twocolumn,showpacs,preprintnumbers,amsmath,amssymb]{revtex4}

\usepackage{graphicx}
\usepackage{dcolumn}
\usepackage{bm}
\usepackage{color}

\begin{document}

\preprint{}

\title{Numerical solution of the $t$-$J$ model with random exchange couplings\\ in $d=\infty$ dimensions}

\author{Junya Otsuki$^{1,2}$}
\author{Dieter Vollhardt$^1$}
\affiliation{%
$^1$Theoretical Physics III, Center for Electronic Correlations and Magnetism,\\
Institute of Physics, University of Augsburg, D-86135 Augsburg, Germany\\
$^2$Department of Physics, Tohoku University, Sendai 980-8578, Japan
}%

\date{\today}

\begin{abstract}
To explore the nature of the metallic state near the transition to a Mott insulator
we investigate the $t$-$J$ model with random exchange interaction in $d=\infty$ dimensions.
A numerically exact solution is obtained by an extension of
the continuous-time quantum Monte Carlo (CT-QMC) method
to the case of a vector bosonic field coupled to a local spin.
We show that the paramagnetic solution near the Mott insulator
describes an incoherent metal with a residual moment,
and that single-particle excitations produce
an additional band, which is separated from the Mott-Hubbard band.
\end{abstract}


\maketitle


The extraordinary properties of high-$T_c$ cuprates are closely related to those of doped Mott insulators \cite{Imada98,Lee06}.
In both systems strong electronic correlations play a key role.
To understand their influence, fundamental electronic correlation models such as the Hubbard and the $t$-$J$ model have been studied intensively \cite{Lee06}.
In spite of their apparent simplicity these quantum mechanical many-particle models can only be solved approximately in dimensions $d=2,3$. Thus the full range of physical phenomena described by the Hubbard or the $t$-$J$ model is not yet understood,
implying that their investigation still
leads to unexpected, and often peculiar results.
For example, it has been pointed that
the Fermi-surface volume of the two-dimensional $t$-$J$ model is inconsistent with the Luttinger-Ward theorem \cite{Putikka98, Phillips06, Kokalj07, Sakai09}.
Indeed, by applying the Schwinger method to the $t$-$J$ model Shastry \cite{Shastry10} recently found a class of solutions which show precisely such a behavior.
In this situation it is desirable to obtain reliable conclusions about those correlation models
at least in certain solvable non-trivial limits. In the case of doped Mott insulators ``non-trivial'' means that characteristic features such as a strong, but screened Coulomb repulsion and the presence of local spin fluctuations in real space are retained.
Namely, the screened Coulomb interaction is responsible for the Mott metal-insulator transition (MIT), and local spin fluctuations affect the quasiparticles by making the self-energy frequency dependent.

The Mott MIT can be described by the dynamical mean-field theory (DMFT) \cite{Metzner89,Georges96}. The DMFT provides an exact, non-trivial solution  of electronic lattice models with a local interaction such as the Hubbard model,  in $d=\infty$.  It can be derived by mapping the original quantum lattice problem onto an effective quantum impurity coupled self-consistently to a dynamical fermionic mean-field (``bath'') \cite{Georges96}.
However, due to the local nature of the DMFT \emph{inter}-site interactions are reduced to a static mean field. This implies that in the strong-coupling limit of the Hubbard model, which corresponds to the Heisenberg model in the case of half-filling and to the $t$-$J$ model in the doped case, non-local spin fluctuations are missing. At the same time it is known from the investigation of spin models in the context of spin-glass problems,
that when the spin coupling $J_{ij}$ is taken to be \emph{random} non-local spin fluctuations survive even in $d=\infty$, while the static mean field averages out \cite{Sherrington-Kirkpatrick75, Thouless77,Bray-Moore80, Grempel98, Georges00a,Georges00b,Sachdev-Ye93}.
In this case the self-consistency equations correspond to those of an effective impurity problem coupled to a bosonic bath.
In particular, the random-coupling Heisenberg model has a very remarkable property: as shown by Sachdev and Ye \cite{Sachdev-Ye93} for SU($M$) spins with $M=\infty$ its dynamical magnetic susceptibility displays marginal Fermi liquid  behavior, which was proposed in the phenomenological theory for the cuprate superconductors \cite{Varma89}.
This suggests a relation of the random-coupling Heisenberg model to the paramagnetic state of the cuprates.
Indeed, spin-glass behavior was observed in La$_{2-x}$Sr$_{x}$CuO$_4$ in the low-doping regime $x=0.04$ \cite{Chou95,footnote-random-coupling}.

The effect of doping on the random-coupling Heisenberg model was studied in detail by Parcollet and Georges \cite{Parcollet99} in terms of the $t$-$J$ model with random couplings $J_{ij}$ in the limit $M=\infty$.
This model also has a non-trivial $d=\infty$ limit and describes a doped Mott insulator without antiferromagnetic spin fluctuations.
In particular, these authors calculated the coherence scale of quasiparticles and discussed the properties of an incoherent metallic state near half-filling.

In this Letter we present a numerically exact solution of the random coupling $t$-$J$ model for the realistic number of spin components, $M=2$.
This is made possible by an extension of the continuous-time quantum Monte-Carlo (CT-QMC) method to a local spin coupled to a bosonic field.
We compute the quasi-particle energy scale as a function of doping and determine the spectrum of the incoherent metal near the Mott insulating state.

{\em The random coupling $t$-$J$ model in $d=\infty$.---}
To obtain a non-trivial $d=\infty$ limit (in the following we use the coordination number $Z=\infty$ instead) where non-local spin fluctuations are retained,
the coupling constants $J_{ij}$ should be scaled as $J_{ij}=J^*_{ij}/\sqrt{Z}$,
with $J^*_{ij}={\rm const.}$ \cite{Parcollet99, Smith-Si00, Haule-tj},
in complete analogy to the scaling of the hopping amplitude \cite{Metzner89, Georges96}.
However, since the static molecular field is proportional to $ZJ$,
the above scaling
leads to a divergence of the transition temperature for magnetic long-range order at half-filling,
i.e., the paramagnetic state is unstable against an infinitesimally small external field for all finite temperatures.
This problem does not occur when the coupling constants $J_{ij}$ are random variables,
since then $\sum_{j}J_{ij}=0$ but $\sum_{j}J^2_{ij}\neq0$,
implying that the static molecular field due to the surrounding sites averages to zero \cite{Thouless77}.
Doping of the random coupling Heisenberg model then leads to the
random coupling  $t$-$J$ model \cite{Parcollet99}
\begin{align}
H = -\frac{t}{\sqrt{Z}} \sum_{\langle ij \rangle \sigma}
 \tilde{c}_{i\sigma}^{\dag} \tilde{c}_{j\sigma}
- \frac{1}{2} \sum_{\langle ij \rangle}
 \frac{J_{ij}}{\sqrt{Z}} \bm{S}_i \cdot \bm{S}_j,
\label{eq:hamil}
\end{align}
where $\tilde{c}_{i\sigma} = c_{i\sigma} (1-n_{i-\sigma})$,
$\bm{S}_i = (1/2) \sum_{\sigma \sigma'} c_{i\sigma}^{\dag} \bm{\sigma}_{\sigma \sigma'} c_{i\sigma'}$.
Here the summation is taken over nearest-neighbor sites,
and we consider the Bethe lattice with infinite connectivity ($Z=\infty$) \cite{footnote-hamil}.
The exchange couplings $J_{ij}$ are randomly distributed according to the probability distribution
\begin{align}
P(J_{ij}) \propto \exp(-J_{ij}^2/2J^2).
\label{eq:prob}
\end{align}

Neglecting spin-glass order,
the model defined by (\ref{eq:hamil}) with $Z=\infty$ reduces to an effective impurity model corresponding to the action \cite{Parcollet99}
\begin{align}
S_{\rm imp} = \int d\tau d\tau'
\Bigg\{ -\sum_{\sigma} f_{\sigma}^{\dag}(\tau) {\cal G}^{-1}(\tau - \tau') f_{\sigma}(\tau')
\nonumber \\
-\frac{1}{2} \bm{S}_f(\tau) \cdot {\cal J}(\tau-\tau') \bm{S}_f(\tau') \Bigg\}
+ \int d\tau U n_{f\uparrow}(\tau) n_{f\downarrow}(\tau),
\label{eq:S_imp}
\end{align}
where the repulsion $U$ is taken to be infinite to exclude double occupation.
The local electron is denoted by the Grassmann number $f_{\sigma}$,
and we introduced
$n_{f\sigma} = f_{\sigma}^{\dag} f_{\sigma}$ and
$\bm{S}_f = (1/2) \sum_{\sigma \sigma'} f_{\sigma}^{\dag} \bm{\sigma}_{\sigma \sigma'} f_{\sigma'}$.
The local propagator ${\cal G}$ and the local time-dependent exchange interaction ${\cal J}$ are
determined by the self-consistency conditions \cite{Parcollet99}
\begin{align}
\label{eq:self-consistent1}
{\cal G}^{-1}(i\omega_n) &= i\omega_n + \mu - t^2 G_{\rm imp}(i\omega_n),\\
\label{eq:self-consistent2}
{\cal J}(i\nu_n) &= J^2 \chi_{\rm imp}(i\nu_n),
\end{align}
where
$G_{\rm imp}$ is the single-particle Green function,
and $\chi_{\rm imp}$ is the spin susceptibility,
both evaluated in the effective impurity model.
Furthermore, $\omega_n$ and $\nu_n$ are fermionic and bosonic Matsubara frequencies, respectively.
The self-consistency condition (\ref{eq:self-consistent2}) corresponds to
that of the extended DMFT \cite{Smith-Si00, Haule-tj, Sun-Kotliar02, Kuramoto-Fukushima98}
with semi-circular density of couplings \cite{footnote-edmft, footnote-scaling}.

The first and second term in $S_{\rm imp}$ can be represented by
fermionic and vector bosonic baths, respectively,
which we express by the operators
$a_{\bm{k}\sigma}$ and $\bm{b}_{\bm{q}}$.
The corresponding Hamiltonian is written as
\begin{align}
\label{eq:H_imp}
H_{\rm imp}
&= -\mu n_f + U n_{f\uparrow} n_{f\downarrow}
+ \sum_{\bm{k} \sigma} \epsilon_{\bm{k}} a_{\bm{k} \sigma}^{\dag} a_{\bm{k} \sigma}
+ \sum_{\bm{q}} \omega_{\bm{q}} \bm{b}_{\bm{q}}^{\dag} \cdot \bm{b}_{\bm{q}}
\nonumber
\\
&+ V \sum_{\sigma} ( f_{\sigma}^{\dag} a_{\sigma} + a_{\sigma}^{\dag} f_{\sigma} )
+ g \bm{S}_f \cdot (\bm{b} + \bm{b}^{\dag}),
\end{align}
where
$a_{\sigma} = N^{-1/2} \sum_{\bm{k}} a_{\bm{k} \sigma}$ and
$\bm{b} = N^{-1/2} \sum_{\bm{q}} \bm{b}_{\bm{q}}$,
with $N$ as the number of sites.
The quantities entering in
$H_{\rm imp}$ are connected with those in $S_{\rm imp}$ by the relations
\begin{align}
\label{eq:bath-f}
\Delta(i\omega_n)
&=\frac{V^2}{N} \sum_{\bm{k}} \frac{1}{i\omega_n - \epsilon_{\bm{k}}},
\\
\label{eq:bath-b}
{\cal J}(i\nu_n)
&= \frac{g^2}{N} \sum_{\bm{q}} \frac{2\omega_{\bm{q}}}{\nu_n^2 + \omega_{\bm{q}}^2},
\end{align}
with
${\cal G}^{-1}(i\omega_n) = i\omega_n + \mu - \Delta(i\omega_n)$.

{\em Spin-Boson Coupling in CT-QMC.---}
We solve the effective impurity model (\ref{eq:H_imp})
using the hybridization-expansion solver of the CT-QMC \cite{Gull11}.
An algorithm for the inclusion of a bosonic bath into the CT-QMC was formulated by Werner and Millis in the case of impurity models with electron-phonon coupling \cite{Werner07,Werner10}.
They used the so-called Lang-Firsov transformation to eliminate the coupling term
and thereby arrived at an efficient treatment of phonons.
An \emph{exchange} coupling
cannot be eliminated in this way, since
the spin operator $\bm{S}_f$ has three components
which do not commute.
For that reason we treat the spin-boson coupling by a stochastic method.
Namely, we perform a double expansion in terms of the hybridization and the spin-boson coupling,
and evaluate the series by Monte Carlo sampling \cite{Otsuki-SB}.

{\em Results.---}
We now present the result of the model~(\ref{eq:hamil})
as a function of the dimensionless coupling strength $J/t$, the particle density $n$, and the temperature $T$.
In the following, energies are measured in units of $W=2t=1$,
where $W$ is the half-width of the density of states of the non-interacting model.


\begin{figure}[tb]
	\begin{center}
	\includegraphics[width=0.8\linewidth]{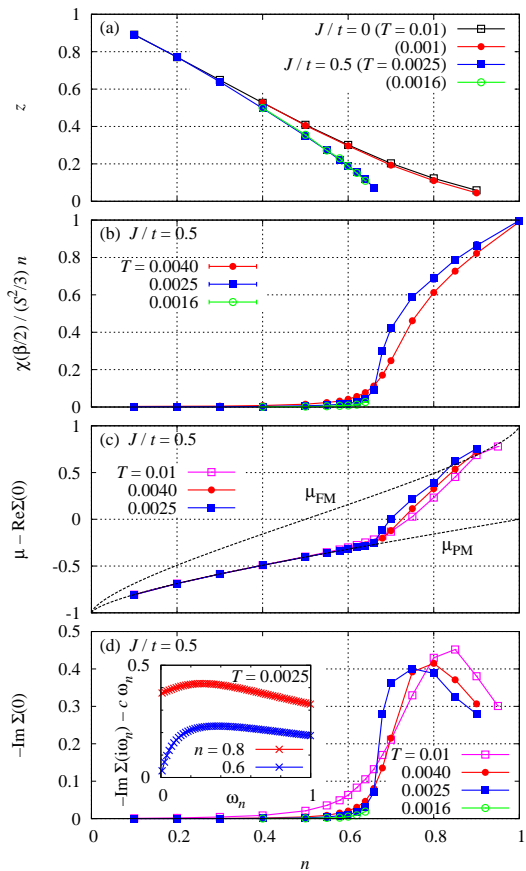}
	\end{center}
	\caption{Four physical quantities characterizing the paramagnetic state are plotted as a function of $n$:
	(a) the renormalization factor $z$,
	(b) the residual moment $\chi(\beta/2)$ per electron normalized by $S^2/3=12$,
	(c) the effective chemical potential $\mu_{\rm eff} \equiv \mu - {\rm Re}\Sigma(0)$, and
	(d) the scattering rate $-\text{Im} \Sigma(0^+)$.
	The dashed curves in (c) show the chemical potential of the paramagnetic and polarized state, $\mu_{\rm PM}$ and $\mu_{\rm FM}$, for a noninteracting system.
	The inset in (d) shows $-\text{Im} \Sigma(i\omega_n)-c\omega_n$ with $c=n/(2-n)$ \cite{footnote-sigma}.}
	\label{fig:n-dep}
\end{figure}

The quasiparticle renormalization factor
$z = [1-{\rm Im}\Sigma(i\omega_0)/\omega_0]^{-1}$
is shown in Fig.~\ref{fig:n-dep}(a)
as a function of density $n$.
While for $J=0$, $z$ tends to zero at $n =1$,
for $J/t=0.5$ it approaches zero at $n \simeq 0.66 \equiv n_{0}$ with increasing slope.
This behavior is contrast to the result in $M = \infty$,
where $z$ vanishes at $n=1$ regardless of the value of $J$ \cite{Parcollet99}.
In the realistic case $M=2$ the system hence appears to be a non-Fermi liquid
for $n \gtrsim n_{0}$.

To identify the type of electronic state which is stable for $n \gtrsim n_{0}$
we calculate the residual moment defined by $\chi(\tau=\beta/2)$,
which is equal to $T\chi$ in the limit $T \to 0$, but which approaches the $T=0$ value faster than $T\chi$.
The residual moment per electron is plotted in Fig.~\ref{fig:n-dep}(b).
The value at $n=1$ corresponds to $S^2/3$, which is consistent with previous QMC calculations \cite{Grempel98}.
We see that $\chi(\beta/2)$ increases with decreasing $T$ for $n \gtrsim n_{0}$,
while it decreases for $n \lesssim n_{0}$.
From this data we conclude that the phase
has a residual moment  for $n \gtrsim n_{0}$.


\begin{figure}[tb]
	\begin{center}
	\includegraphics[width=\linewidth]{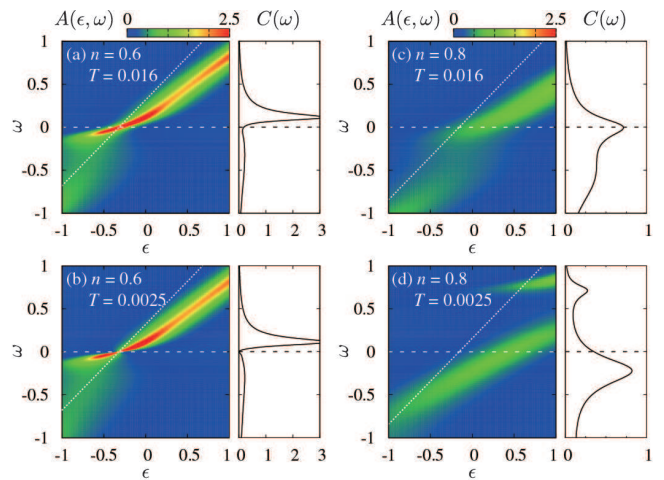}
	\end{center}
	\caption{Intensity plots of the single-particle excitation spectrum $A(\epsilon, \omega)$, and the `atomic spectrum' $C(\omega)$ for $J/t$=0.5, $n$=0.6, 0.8 and $T$=0.0025, 0.016.
	The dotted line shows the energy dispersion of the non-interacting system.}
	\label{fig:spectra}
\end{figure}

Next, we compute the momentum-resolved single-particle excitation spectrum
$A(\epsilon, \omega)$, which is defined in terms of
the local self-energy
$\Sigma(i\omega_n)={\cal G}^{-1}(i\omega_n) - G_{\rm imp}^{-1}(i\omega_n)$
by
\begin{align}
A(\epsilon, \omega)
= -\frac{1}{\pi} {\rm Im} \frac{1}{\omega^+ +\mu -\epsilon - \Sigma(\omega^+) }.
\end{align}
Here $\epsilon \equiv \epsilon_{\bm{k}}$, with $\epsilon [-W, W]$,  is the dispersion in $d=\infty$
which can be used to parameterize the momentum dependence.
We performed the analytical continuation to $\omega^+ = \omega + i0$ by a Pad\'{e} approximation \cite{footnote-Pade}.
The results are shown in Fig.~\ref{fig:spectra}.
For $n=0.6$, i.e., in the Fermi liquid regime
[Figs.~\ref{fig:spectra}(a) and \ref{fig:spectra}(b)],
quasiparticle excitations are seen to occur around the Fermi energy.
Fig.~\ref{fig:spectra}(c) and (d) show the spectra
for $n=0.8$, i.e., for the state with a residual-moment. At the higher temperature the dispersion is similar to that in the Fermi-liquid regime,
but the spectral features are less sharp than those for $n=0.6$.
In fact, they do not become sharper even for the lowest temperature of our calculation [Fig.~\ref{fig:spectra}(d)].
We conclude from these results,
that the phase close to half-filling corresponds to an \emph{incoherent} metallic state with a residual moment.

The spectrum shown in Fig.~\ref{fig:spectra}(d) has very remarkable features.
Firstly, the broad spectrum crosses the Fermi level at the `momentum' $\epsilon$ which is larger than its value for the non-interacting case.
Secondly, there is an additional band above the broad spectrum.
We now discuss these features in more detail.
The effective chemical potential
$\mu_{\rm eff} = \mu - {\rm Re}\Sigma(0)$,
which is related to the Fermi-surface volume if ${\rm Im}\Sigma(0)=0$,
is shown in Fig.~\ref{fig:n-dep}(c)
together with the chemical potentials of the paramagnetic and polarized state,
$\mu_{\rm PM}$ and $\mu_{\rm FM}$,
in the non-interacting system ($U=J=0$).
For $n \lesssim n_{0}$, $\mu_{\rm eff}$ agrees with $\mu_{\rm PM}$,
indicating that the Luttinger theorem is satisfied.
On the other hand, it starts to deviate from $\mu_{\rm PM}$ at $n \simeq n_{0}$
and approaches $\mu_{\rm FM}$.
We note that this deviation does not imply a violation of the Luttinger theorem,
since there is no discontinuity in the momentum distribution function in this regime.
Namely, for densities $n \gtrsim n_0$  the scattering rate ${-\rm Im} \Sigma(0^+)$  does not tend to zero for temperatures down to $T=0.0025$ (Fig.~\ref{fig:n-dep}(d)).
This is also 
 explicitly seen in the $\omega_n$ dependence of ${\rm Im} \Sigma(i\omega_n)$ (inset of Fig.~\ref{fig:n-dep}(d)).

The origin of the additional band observed in Fig.~\ref{fig:spectra}(d) can be explained as follows.
The weights of the upper and lower Hubbard band vary according to $n/2$.
Hence, upon hole-doping some weight is transfered from the Hubbard bands
to energies just above the lower band edge,
leading to an additional spectral weight $\delta=1-n$ \cite{Meinders93}.
This additional spectrum is due to the hole dynamics.
In Fig.~\ref{fig:spectra}(d) the hole spectrum is separated from the lower Hubbard band,
while in the coherent Fermi liquid regime the two mix.
Hence, the appearance of the additional band in Fig.~\ref{fig:spectra}(d)
indicates the incoherence of the holes.
This is also clearly expressed by the `atomic spectrum'
$C(\omega)=-(1/\pi) \text{Im}[\omega^+ + \mu - \Sigma(\omega^+)]^{-1}$
in Fig.~\ref{fig:spectra}.
For $n=0.6$ $C(\omega)$ shows a single sharp peak corresponding to the coherent band, while for $n=0.8$ it consists of two slightly broader peaks corresponding to the two-band structure.
A similar single-particle spectrum was observed for a single hole in the two-dimensional $t$-$J$ model \cite{Dagotto94}.

\begin{figure}[tb]
	\begin{center}
	\includegraphics[width=0.9\linewidth]{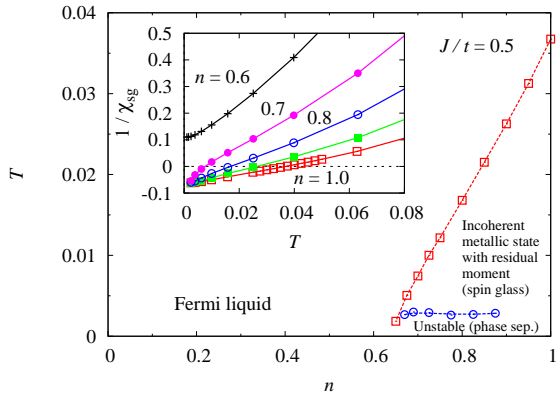}
	\end{center}
	\caption{Temperature vs. density phase diagram of the random coupling $t$-$J$ model in $d=\infty$ for $J/t=0.5$. The blue dashed curve denotes the boundary of a phase separated region.
	The red dashed curve indicates the transition to a spin-glass state.
	The inset shows the inverse of the spin-glass susceptibility $\chi_{\rm sg}$ as a function of $T$.}
	\label{fig:phase}
\end{figure}

At this point a comment on the possible ground state near half-filling is in order.
The incoherent metallic state has a residual entropy,
and therefore
an instability is expected to take place to lift the degeneracy.
Indeed, in our calculation with a paramagnetic bath
we found a divergence of the charge compressibility
$\partial n/\partial \mu$ below $T \lesssim 0.003$,
indicating that
the paramagnetic solution is unstable against phase separation.
The temperature at which $\partial \mu/\partial n$ changes sign is plotted in Fig.~\ref{fig:phase} (blue dashed curve).
Phase separation has also been found in the one-dimensional $t$-$J$ model for large-$J$ \cite{Ogata91}.
Another possibility is magnetic symmetry breaking.
Furthermore, it is known that at half-filling a spin-glass transition,
i.e., the breaking of the replica symmetry, occurs \cite{Sherrington-Kirkpatrick75, Thouless77, Bray-Moore80, Grempel98, Georges00a,Georges00b},
while long-range magnetic order is suppressed due to the random distribution of the exchange interaction.
To estimate the spin-glass transition temperature $T_{\rm sg}$,
we evaluate the spin-glass susceptibility $\chi_{\rm sg}$
using the expression
\begin{align}
\chi_{\rm sg} = \chi_{\rm imp}^2 / (1-J^2 \chi_{\rm imp}^2),
\end{align}
derived at half-filling \cite{Georges00b}.
In the inset of Fig.~\ref{fig:phase} the temperature dependence of $1/\chi_{\rm sg}$ is shown.
The transition temperature $T_{\rm sg}$ itself is plotted in Fig.~\ref{fig:phase} (red dashed curve).
At $n=1$, we obtain
$T_{\rm sg}/J \approx 0.147$,
which is consistent with the result of Ref.~\cite{Grempel98}.
Upon doping, $T_{\rm sg}$ decreases monotonously and reduces to zero at $n \simeq 0.635$, a value which is close to $n_{0}$.
Therefore the incoherent metallic state is actually located
in the region where the spin-glass phase can be expected to be stable.
That is, the incoherent metallic state may be stabilized
if the spin-glass transition is suppressed.
Nevertheless, the peculiar spectrum in this regime is still physically meaningful,
since the divergence of $\chi_{\rm sg}$ does not affect the self-consistency equations.
This is the same as in the case of the paramagnetic DMFT solution of the Hubbard model,
which is found in the region where actually the antiferromagnetic phase is stable \cite{Georges96}.

In summary,
we presented the exact numerical solution of the $t$-$J$ model with random exchange couplings in $d=\infty$.
Near half-filling the solution  corresponds to an incoherent metal
with a residual moment.
The single-particle excitations not only lead to a broad spectrum crossing the Fermi level,
but to an additional band at higher energies.
Because of the residual moment, the effective chemical potential is located close to the Fermi level of the polarized non-interacting system.
The additional band is observed only in the
non-Fermi liquid regime,
and is a signature of the incoherence of the holes.

Finally, we comment on the $t$-$J$ model with {\em non-random} couplings in $d=2, 3$.
Eqs.~(\ref{eq:self-consistent1}) and (\ref{eq:self-consistent2})
may be regarded as a single-site approximation for models in $Z<\infty$ \cite{footnote-scaling}.
Indeed, we obtained preliminary results for the same quantities as shown in Fig.~\ref{fig:n-dep} in $d=2$ and 3 which indicate that the paramagnetic solution
near half-filling is again an incoherent metal with a residual moment.
The case $d<\infty$,
including symmetry broken solutions,
e.g., the antiferromagnetic or the spin-glass state, will be investigated in the future.

We thank A. Georges, Y. Kuramoto, T. Pruschke, and G. Zar\'and for useful discussions.
Support of one of us (J.O.) by a JSPS Postdoctoral Fellowship for Research Abroad,
and by the Deutsche Forschungsgemeinschaft through TRR 80 is gratefully acknowledged.


\begin{thebibliography}{99}
\bibitem{footnote-random-coupling} Subsequent studies of the SU($M$) random coupling Heisenberg model within DMFT for $M=2$ by quantum Monte Carlo \cite{Grempel98} and by a $1/M$ expansion \cite{Georges00a,Georges00b} did not find marginal Fermi-liquid behavior.
\bibitem{footnote-hamil}
The Hamiltonian (\ref{eq:hamil}) in $Z=\infty$ is equivalent to a $t$-$J$ model on a fully connected lattice
with random hopping and random exchange interactions.
This correspondence was demonstrated for the Ising model in Ref.~\cite{Georges90}.
\bibitem{footnote-edmft}
This correspondence follows from the fact
that the eigenvalues of the random matrix $J_{ij}$ with matrix elements $J_{ij}$ distributed
according to (\ref{eq:prob}), have
a semi-circular distribution with half-width $2J$ \cite{Thouless77}.
\bibitem{footnote-scaling}
We note that when Eq. (\ref{eq:self-consistent2}) is applied to models without randomness,
the coupling constants should be scaled as $J_{ij} \sim 1/Z$ to keep the magnetic transition temperature finite.
In this case, the equation incorporates all the terms of order $1/Z$ as well as higher-order terms within a single-site approximation \cite{Kuramoto-Fukushima98}.
\bibitem{footnote-sigma}
At $U=\infty$ the self-energy $\Sigma(i\omega_n)$ 
has the high-frequency behavior $\Sigma(i\omega_n) \sim (1-1/a) i\omega_n$, which originates in $G(i\omega_n) \sim a/i\omega_n$ with $a=1-n/2$. We subtracted this term in the plot.
\bibitem{footnote-Pade} Here a Pad\'{e} approximation is possible because the CT-QMC data have very little statistical noise.
\end{thebibliography}
\end{document}